\documentclass[iop]{emulateapj}
\usepackage{amsmath,amssymb,graphicx,soul,color}
\usepackage{epstopdf}
\usepackage{soul}
\usepackage{float}
\usepackage{booktabs}

\usepackage{natbib}
\bibpunct{(}{)}{;}{a}{,}{,}
\begin{document}

\title{Gamma-Ray Burst Flares: X-ray Flaring. II.}

\slugcomment{Submitted to ApJ}

\author{C. A. Swenson\altaffilmark{1}, P. W. A. Roming\altaffilmark{2,1}}

\email{cswenson@astro.psu.edu}
\altaffiltext{1} {Pennsylvania State Univ., 525 Davey Lab, University Park, PA 16802, USA}
\altaffiltext{2}{Southwest Research Institute, 6220 Culebra Road, San Antonio, TX 78238, USA}

\begin{abstract}

We present a catalog of 497 flaring periods found in gamma-ray burst (GRB) light curves taken from the online XRT GRB Catalogue.  We analyzed 680 individual light curves using a flare detection method developed and used on our UV/optical GRB Flare Catalog.  The method makes use of the Bayesian Information Criterion to analyze the residuals of fitted GRB light curves and statistically determines the optimal fit to the light curve residuals in attempt to identify any additional features.  These features, which we classify as flares, are identified by iteratively adding additional `breaks' to the light curve.  We find evidence of flaring in 310 of the analyzed light curves.  For those light curves with flares, we find an average number of $\sim$1.5 flares per GRB.  As with the UV/optical, flaring in our sample is generally confined to the first 1000 s of the afterglow, but can be detected to beyond $10^{5}$ s.  Only $\sim$50\% of the detected flares follow the `classical' definition of $\Delta t/t \ll 1$, with many of the largest flares exhibiting $\Delta t/t \sim 1$ and some exceeding this value.
\end{abstract}


\section{Introduction}

The \emph{Swift} \citep{Gehrels2004} mission has revolutionized the study of Gamma-ray Burst (GRB) afterglows due to its rapid response time and automated GRB response algorithms.  \emph{Swift} is comprised of three instruments that work together and all contribute their unique capabilities to this the study of GRBs.  The Burst Alert Telescope \citep[BAT; ][]{Barthelmy2005} first detects the GRB and causes the satellite to perform an autonomous slew to the GRB position, generally within $\sim$100 seconds of the GRB trigger.  The X-ray Telescope \citep[XRT; ][]{Burrows2005} and UV/Optical Telescope \citep[UVOT; ][]{Roming2000,Roming2004,Roming2005} then begin an automated sequence of observations designed to localize the position of the GRB to less than an arcsecond and follow the decay of the afterglow.  \emph{Swift} has triggered on and localized an X-ray counterpart for over 700 GRBs, increasing the number of afterglow localizations by approximately an order of magnitude from the pre-\emph{Swift} era.  The rapid localizations and follow-up provided by \emph{Swift} has resulted in several exciting new discoveries about the properties of the GRB afterglow, including the ``canonical" X-ray light curve \citep{Nousek2006}, which is observed in a number of GRBs (e.g. \citealt{Hill2006}, \citealt{Evans2009}).  Also discover early in the \emph{Swift} mission was the presence of X-ray flares in the early afterglow (e.g. \citealt{Burrows2005a}, \citealt{Romano2006}).

Flares in GRB afterglows had been seen previous to their observation by the XRT (e.g \citealt{Piro1998,Piro2005}), but only in three X-ray afterglows.  The XRT observations have shown that flares are seen in phases of the canonical X-ray light curve and quite common, appearing in approximately 50\% of the XRT afterglows \citep{OBrien2006}.  These flares are observed as superimposed excesses deviating from the underlying light curve.

Through the study of individual flares in GRBs 050406 \citep{Romano2006}, 050502B \citep{Falcone2006}, 050713A \citep{Morris2007}, 050724 \citep{Campana2006} and 050904 \citep{Cusumano2007} it has been shown that X-ray flares are observed in both long and short GRBs, appear to come from a distinctly different emission mechanism than the underlying afterglow, and can be observed out to beyond $10^{5}$ seconds from the initial GRB trigger (e.g. \citealt{Swenson2010}).  These studies also point toward a likely internal shock source for the flares, though the actual source of the flares still remains in question and may be caused by one of many different mechanisms including instabilities in the ejecta, stored electromagnetic energy or collision with the extrastellar medium \citep{Zhang2006a}.

Studying large numbers of flares from several different bursts, and analzying their bulk properties allows us to try and better constrain the physical process by which flares are created.  Several previous studies have been performed on XRT light curves, analyzing groups of X-ray flares.  The earliest studies by \citet{Falcone2007} and \citet{Chincarini2007} examined 33 flares in the first 110 GRBs observed by \emph{Swift}.  These studies showed that late-time internal shocks were necessary to explain the 10 of the observed flares and that some sort of central engie activity was the preferred method for a majority of the flares.  Follow-up studies performed by \citep{Chincarini2009,Chincarini2010} and \citep{Margutti2010} showed that X-ray flaring may have some correlation with the GRB prompt emission and showed that flares evolve over time, and were likely caused by late-time internal dissipation processes.

\citet{Morris2008} did incorporate the BAT, XRT and UVOT data for the flare sample used by \citet{Falcone2007} and \citet{Chincarini2007} and showed that wheras the afterglow could be fit by a simple absorbed power law, the SED of the flaring periods could not.  \citep{Roming2006a} attempted to perform a study analyzing flares in the UV/optical but was severely limited due to the low significance of UV/optical flares compared to X-ray flares.

Realizing the need for more detected flares in the UV/optical, in our first paper \citep{Swenson2013a} we presented a catalog of flares found in the UV/optical from the collection of light curves presented in the Second \emph{Swift} Ultraviolet/Optical Telescope GRB Afterglow Catalog, an expansion on the First \emph{Swift} Ultraviolet/Optical Telescope GRB after Catalog \citep{Roming2009}. These flares were found using a new algorithm developed specifically for the purpose of performing a blind, systematic search for flares in GRB afterglows.  This search resulted in the discovery of 119 potential flaring periods in 68 GRB afterglows, many of which were previously undetected.  This study showed that flares in the UV/optical are much more common than has previously been thought.

As mentioned previously, many of the studies on X-ray flares were limited in their reach due to limitations in their data sets.  \citet{Falcone2007} and \citet{Chincarini2007} were limited by the time that \emph{Swift} had been operating, \citet{Chincarini2009} limited their data set to GRBs with redshift measurements to study the actual energetics, \citet{Chincarini2010} limited their study to only flares found within the first 1000 seconds of the GRB afterglow, and \citet{Margutti2010} used only a sample of 9 exceptionally bright X-ray flares.  There has yet to be a blind, systematic search for X-ray flares that is not somehow limited in scope.

The precise nature of the GRB central engine is still largely unknown and many of the previous studies on GRB flares have indicated a likely connection between flaring and the central engine, making the study of GRB flares crucial to our understanding of GRBs.  Having a complimentary X-ray catalog using the algorithm developed in \citet{Swenson2013a} would address the limitations mentioned in the previous X-ray studies and allow for more stringent constraints on the origin of GRB flares through cross-correlation of the X-ray and UV/optical.

In this paper we present the results from a blind, systematic search for flares in XRT light curves.  Using the method described in \citet{Swenson2013a} we have constructed the most complete catalog of X-ray flares to date and provide the temporal details of each flare, including $T_{peak}$, $\Delta t/t$, and the strength of the flare relative to the underlying light curve.  In a forthcoming paper we will perform our cross-correlation analysis of this catalog and our UV/optical flare catalog.

This paper is organized as follows:  In \S 2 we describe our data set as well as our methodology for identifying flares.  We present our catalog of GRB flares observed by the XRT in \S 3, and discuss the implications drawn from the catalog in \S4. 

\section{Methodology}

For the purposes of this study we will use the publicly available XRT light curves from the online Swift-XRT GRB Catalogue \citep{Evans2007,Evans2009}.  We downloaded the light curves for the time period covering January 2005 through December 2012, inclusive, as well as the best fit parameters for each burst.  We calculated the light curve residuals using the best fit parameters and perform our flare finding analysis on these residuals.

Our flare finding analysis follows the same basic methodology set forth in \citet{Swenson2013a}.  We processed the calculated residuals using the breakpoints analysis function \citep{Zeileis2003} within the publicly available \texttt{R} \citep{Rcitation} package \texttt{structchange} \citep{Zeileis2002}.  The breakpoints analysis determines the optimal number of `breakpoints' that are required to best explain any features that may remain in the residuals of the GRB light curve.  This is done by simultaneously minimizing both the residual sum of squares (RSS) and the Bayesian Information Criterion \citep[BIC; ][]{Schwarz1978} over several iterative fits to the light curve.  For the purposes of our analysis we used the guidelines provided by \citet{Kass1995} and require $BIC_{i} - BIC_{min} > 6$ as the criteria for determining the preferred fit to the light curve residuals.

We performed $10,000$ Monte Carlo iterations for each GRB light curve, each time varying the datapoints to account for measurement error and then determining the optimal number of breakpoints and grouping those breakpoints into potential flares.  For each potential flare we identify the following parameters:  $T_{Start}$, $T_{peak}$ and $T_{Stop}$, the start, peak and end times of each flare, respectively, each nominally associated with an individually identified breakpoint.  Due to the higher density of data points, and therefore timing resolution, our determinations of $T_{start}$ and $T_{stop}$ will be more precise than for the UV/optical flares, but we will continue to refer to them as `limits' because there are still instances of poor timing resolution and gaps in the data that prevent us from determining a more accurate breakpoint.  We calculated $\Delta t/t$, defined as ($T_{Stop}$ - $T_{Start}$)/$T_{peak}$, and the peak flux ratio using the measured flux at $T_{peak}$ and an interpolation of the flux of the underlying light curve at the same time.  We also provide a confidence measure, which we define as the fractional number of times that a particular flare was recovered during the $10,000$ simulations.

A few minor changes in the actual processing of the data were required, as opposed to the UV/optical dataset.  Due to the much higher density of data points available in many of the X-ray light curves, as opposed to the relatively sparcely sampled UV/optical light curves, were were forced to limit the number of potential breakpoints identified to 75 per light curve.  By default the analysis iteratively adds additional breakpoints between ever data point in the light curve, beginning with the strongest (i.e. most likely) breakpoint.  This process is computationally intensive and adding an arbitrarily large number of additional breakpoints increases the processing time exponetially.  By limiting the number of breakpoints to 75 we are allowing for a minimum of 25 individual flares per light curve.  Our results presented in this paper show that no burst had more than nine individual flares identified, so the truncation of the analysis had no effect on the end results.

Additionally, due to the number of data points contained in some of the brightest X-ray light curves, the process of iteratively fitting every data point requires a large number of CPU cycles and completing the normal $10,000$ Monte Carlo iterations would have required several years of computional time.  In those cases we limited the number of iterations to $1,000$ Monte Carlo simulations and report our confidence measure as the fraction of times the flare was recovered for those $1,000$ simulations. 

\section{Results}

Here we present the results of our analysis of the 680 XRT GRB light curves taken from the online Swift-XRT GRB Catalogue \citep{Evans2007,Evans2009} spanning January 2005 to December 2012, inclusive.  We detect 497 unique potential flaring periods, for which we can distinguish start and stop times, detected in 324 different light curves.  A number of these identified flares are actually multiple superimposed flares contained within a shared `flaring period'.  Because of the high density of data points in the X-ray light curves, we are able to resolve periods of multiple overlapping flares.  Due to the overlapping, we can not uniquely identify the start or stop of the individual flares within the larger `flaring period'.  We are limited to identifying only the start and stop times of the entire period containing the overlapping flares.  For the sake of simplicity and completeness we will include these flaring periods in our analysis and simply refer to these flaring periods as `flares'.  Table~\ref{tab:Flaretable} provides the following information for each potential flare:  (1) GRB Name, (2) the flare peak time, defined as the data point most often identified as the flare peak during the Monte Carlo simulations, as well as limits on (3) $T_{start}$ and (4) $T_{stop}$, defined as the last and first data points, respectively that are well fit by the underlying light curve.  (5) a limit on $\Delta t/t$ based on the peak time, $T_{start}$ and $T_{stop}$, and (6) the ratio of the peak flux during the flaring period, relative to the flux of the underlying light curve at the same time, using the observed flux at the flare peak time and an interpolation of the flux of the underlying light curve.  The flux ratio is normalized using the flux of the underlying light curve to allow for direct comparison of each flare across all light curves.  Finally, (7) the confidence measure of the detected flare indicating the fractional number of times the flare was recovered during the $10,000$ Monte Carlo simulations.  We have noted the overlapping 'flaring periods' with an `$\ast$' next to the GRB name in Table~\ref{tab:Flaretable}.

\section{Discussion}

Our analysis shows that at least 47\% of the analyzed XRT light curves contain possible flaring episodes.  This percentage is very similar to previous studies (e.g. \citealt{OBrien2006,Chincarini2010}), in spite of our detection of a significantly larger number of total flares and specifically a larger number of small, weak flares.  This may indicate that X-ray GRB afterglows comes in two varieties:  those with flares and those without.

In our analysis of the bulk properties of the detected X-ray flares we have followed the same method used in \cite{Swenson2013a} and divided the flares into three groups:  ``gold'', ``silver'' and ``bronze''.  Our comparisons to uv/optical flares will also come from our analysis found in \cite{Swenson2013a}.

The gold group is defined as those flares with confidence measure greater than $0.7$ and $\Delta t/t \le 0.5$.  This group constitues those flares which satisfy the somewhat ``classical'' definition of a flare in terms of duration and have a good recoverability rate.  This group contains  127 flares.  The silver group allows for longer flares and lower confidence, expanding the parameters to confidence measure greater than 0.6 and $\Delta t/t \le 1.0$.  This group contains 115 flares after excluding overlap from the gold group.  The remaining flares that do not qualify for either the gold or silver are grouped together in the bronze, which contains 255 flares.

\begin{figure}
	\includegraphics[width=0.6\textwidth,angle=-90]{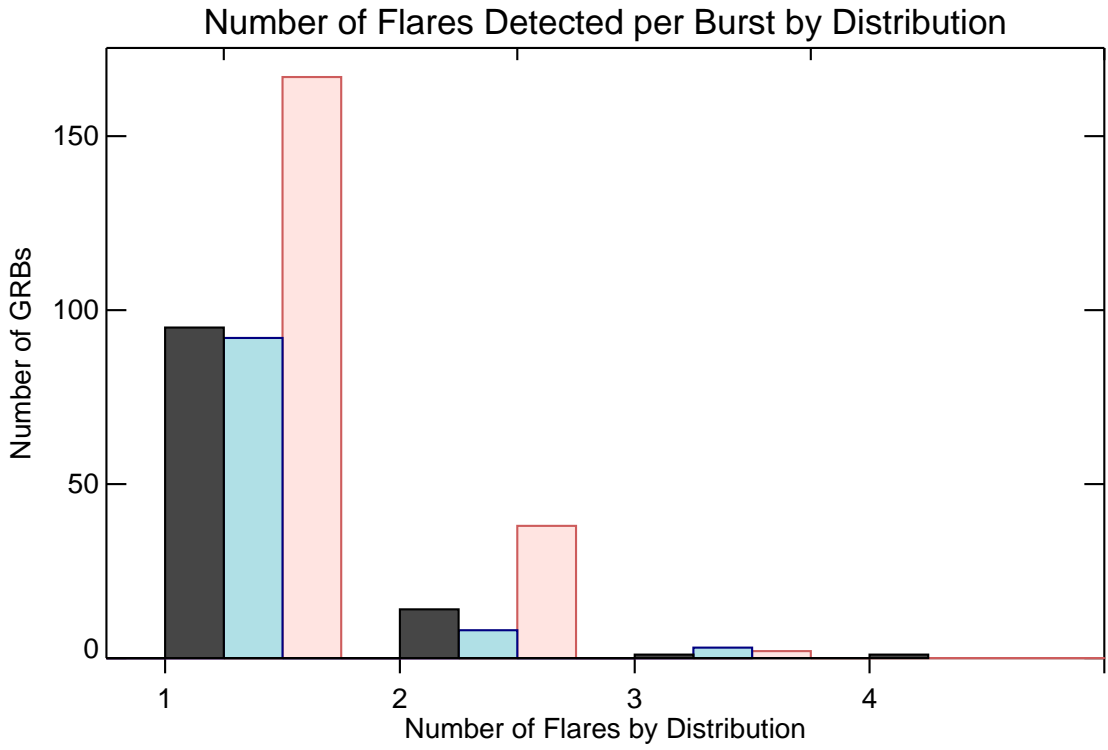}
	\caption{Histogram of the number of detected flares per GRB by group.  The three distributions are the gold (black font), silver (blue font) and bronze (red font) groups described in the text.}
	\label{fig:Number_of_Flares_Histogram}
\end{figure}

Of the 323 X-ray light curves with flares, the average number of flares per GRB is $\sim$1.5.  Figure~\ref{fig:Number_of_Flares_Histogram} shows the distribution of flares per GRB for the gold, silver and bronze groups, shown in black, blue, and red, respectively.  GRB100728A had the most resolved flares of the analyzed bursts, with nine, and five other GRB light curves had five or more flares.

\begin{figure}
	\includegraphics[width=0.6\textwidth,angle=-90]{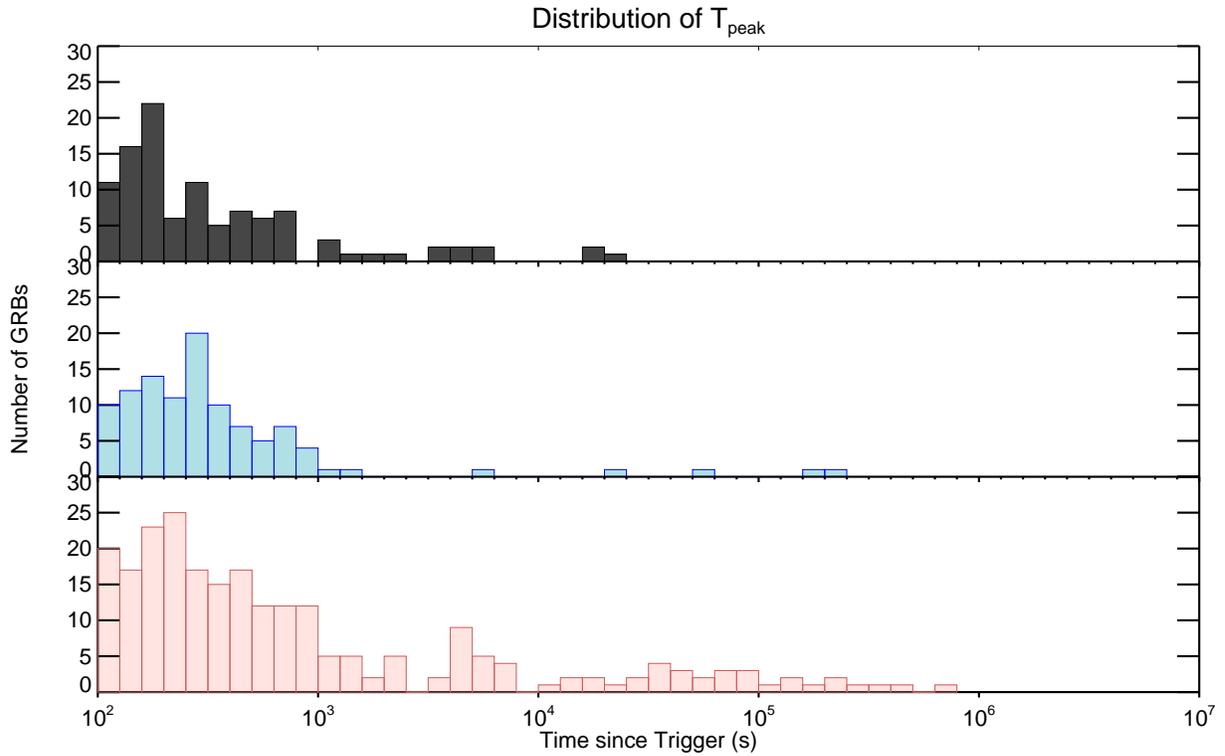}
	\caption{Histogram of the distribution of $T_{peak}$.  The three distributions are the gold (top), silver (middle) and bronze (bottom) distributions described in the text.}
	\label{fig:T_peakHistogram}
\end{figure}

The flare peak times range from between 48 s after the trigger of GRB 110119A to over 400 ks for GRB 090902B.  82\% of all detected flares peaked before 1000 s, nearly matching the percentage seen in the UV/optical light curves.  We suspect that this similarity to the UV/optical flares is not coincidental and that many of these flares may be correlated, or at the very least caused by a similar mechanism that is active during the early stages of the GRB.  This issue will be looked at in depth in our next paper correlating the UV/optical and X-ray flares.  Figure~\ref{fig:T_peakHistogram} shows the distribution of $T_{peak}$ for the three groups of flares.  The grouping of $T_{peak} \le 1000$ s is immediately obvious in all three groups, and all three groups appear to originate from a similar parent distribution peaking between 300 s and 500 s after the trigger.

\begin{figure}
	\includegraphics[width=0.6\textwidth,angle=-90]{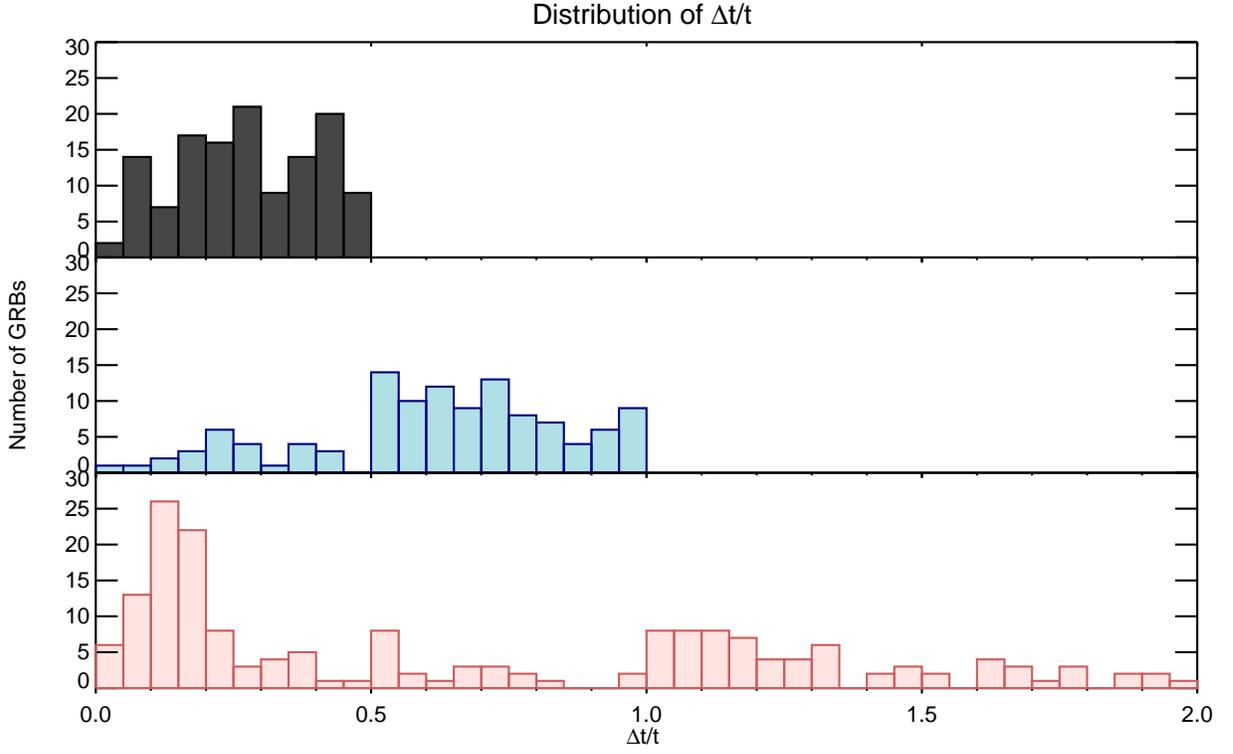}
	\caption{Distribution of $\Delta t/t$, calculated as ($T_{stop} - T_{start})/T_{peak}$, for the detected flares.  The three flares with $\Delta t/t > 2.0$ are omitted for scaling purposes.  The three distributions are the gold (top), silver (middle) and bronze (bottom) distributions described in the text.}
	\label{fig:delta_t_t_Histogram}
\end{figure}

The duration of the flares, recognizing that a number of the $T_{start}$ and $T_{stop}$ values are only limits, vary from $\Delta t/t$ of 0.02 to over 100 (though the extremely large values are due to observing gaps in the data).  Only $\sim$50\% of the flares exhibited $\Delta t/t \le 0.5$, whereas this number was at least 80\% for the UV/optical flares.  This difference between the duration of the X-ray and UV/optical flares may be due to the UV/optical flares being generally fainter than those seen in the X-ray.  If we only see the peak of the flare in the UV/optical, then our measured duration for the flare will be biased relative to the X-ray where we see more of the flare rise and decay.  Figure~\ref{fig:delta_t_t_Histogram} shows the distribution of $\Delta t/t$ for the three groups of flares.  \citet{Ioka2005} showed that it is difficult to achieve rapid variability, defined as $\Delta t/t \le 1$, in the external shock and so an internal shock model has been favored to explain the $\Delta t/t \ll 1$ seen in most flares.  However, Figure~\ref{fig:delta_t_t_Histogram} shows a significant number of possible flares that exhibit $\Delta t/t > 1$.  For this work we are reporting all potential features detected by our flare finding algorithm, and we treat them as potential flares.  It is possible, however, that a portion of our detected features, in particular those with $\Delta t/t \ge 1$, are due to other processes, such as the emergence of the reverse shock, and are not flares.  It is also possible that these are flares caused by processes other than internal shocks.  An interesting relationship between the gold, silver and bronze groups needs to be pointed out when interpreting Figure~\ref{fig:delta_t_t_Histogram}.    There is a continuous distribution of potential flares that is spread across the three groups.  We split the detected flares into three groups based on the prior understanding of flare properties, namely $\Delta t/t \ll 1$, and so created groups based on the combination of flare recoverability and $\Delta t/t$.   Many of the very large and most easily identified potential flares exhibit $\Delta t/t > 0.5$ and some $\Delta t/t > 1.0$.  Because the flares do not meet the criteria for the gold group they spill over into the silver and bronze groups.  This can be seen by the abrupt cut-off, based on our group criterion, in the gold group at $\Delta t/t = 0.5$ and the subsequent continuation of the distribution in the silver group between $0.5 < \Delta t/t \le 1.0$ and the excess tail extending into the bronze group at $\Delta t/t > 1.0$.  These large flares comprise the majority of the silver group, with the remaining flares being distributed at $\Delta t/t <0.5$.  The primary distribution of the bronze flares, removing the extended tail from the gold and silver groups, can be see at $\Delta t/t < 1.0$ and peaking at $\Delta t/t \sim0.1$

\begin{figure}
	\includegraphics[width=0.6\textwidth,angle=-90]{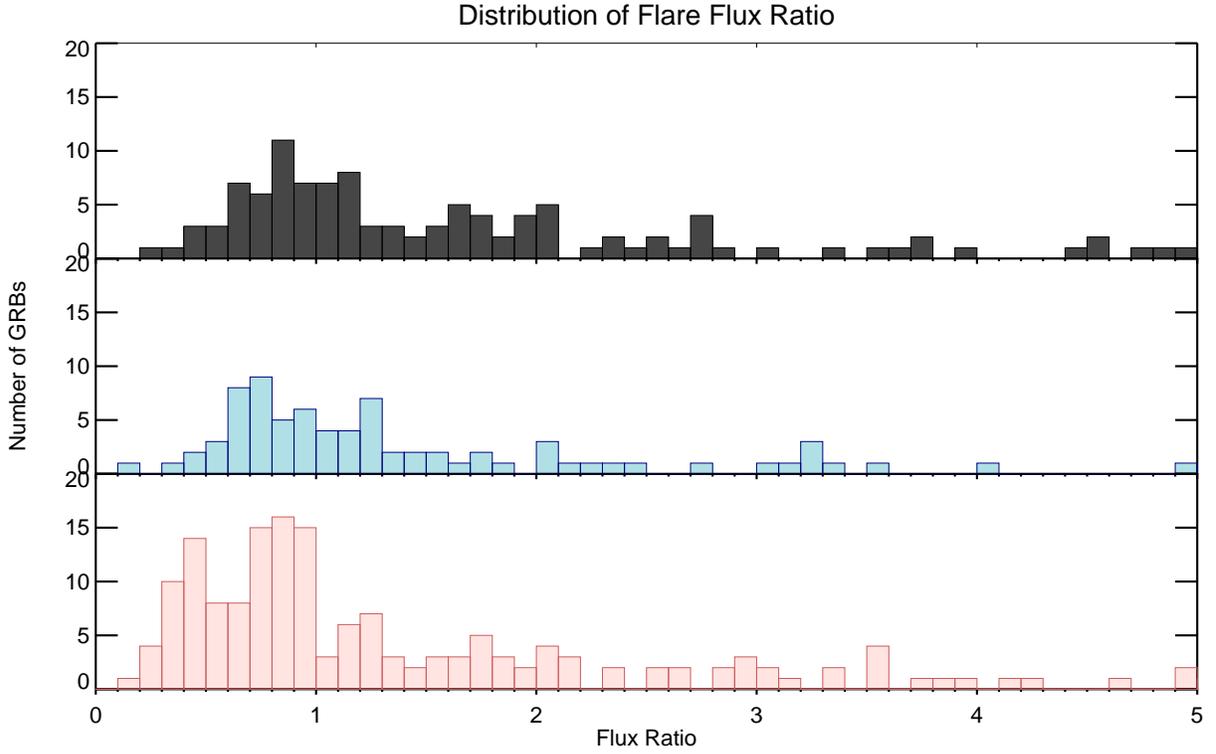}
	\caption{Distribution of flare flux ratio, relative to the underlying light curve.  The flares with flux ratios $> 5$ are omitted for scaling purposes.  The three distributions are the gold (top), silver (middle) and bronze (bottom) distributions described in the text.}
	\label{fig:Flux_Ratio_Histogram}
\end{figure}

The relative strengths of the flares ranges from a minimum flux ratio of 0.1 to a maximum of several thousand.  Figure~\ref{fig:Flux_Ratio_Histogram} shows the distribution of flare flux ratios for the three groups for values $< 5$.  All three groups of flux ratios have long tails that extend into the tens, hundreds, and thousands for the gold, silver, and bronze groups, respectively.  The flux ratios shown in Figure~\ref{fig:Flux_Ratio_Histogram} show the distributions for those smaller, weaker flares that have previously been less studied.  Unlike the UV/optical flares, which had noticeable gaps in the distributions of flux ratios, the X-ray flares show a much more continuous distribution.  The silver group appears to have either a bimodal distribution with peaks at 0.6 and 1.3, or a continuous distribution that is suppressed near a flux ratio of 1.0.  This same suppression also appears in the bronze distribution as a sudden dropoff at flux ratios $> 1.0$, rising again to a peak at 1.3.  The gold group does not exhibit the suppression at flux ratio of 1.0  but is well fit by a distribution centered at $\sim$0.8.  Only a small number (17\%) of UV/optical flares were considered to be strong flares with flux ratios $>2$.  By that same criteria 33\% of X-ray flares are considered large, showing the relative strength of X-ray flares compared to the UV/optical flares.

\section{Conclucsions}

We have analyzed 680 XRT GRB light curves from the online Swift-XRT GRB Catalogue \citep{Evans2007,Evans2009} using the flare detection method introduced in \cite{Swenson2013a}.  We detect the presence of 497 unique potential flaring periods, many of them previously unreported.  We plan to perform a cross- correlation analysis of the UV/optical flares provided in \cite{Swenson2013a} and the X-ray flares reported in this work.  By using the multi-wavelength flare information from these two catalogs we will be able to better constrain the properties of GRB flares and better understand their origin.

\acknowledgements
We thank Eric Feigelson for his help in pointing us toward the use of the Bayesian Information Criterion.
This work made use of the data supplied by the UK Swift Science Data Centre at the University of Leicester.

\clearpage
\LongTables
\begin{deluxetable*}{clrrrrrr}
\tabletypesize{\scriptsize}
\tablecaption{X-ray Flares}
\tablewidth{0pt}
\tablecolumns{8}
\tablehead{
\colhead{Flaring Period} &
\colhead{Source Name} &
\colhead{$T_{peak}$*} &
\colhead{$T_{start}$ lower limit*} &
\colhead{$T_{stop}$ upper limit*} &
\colhead{$\Delta$t/t} &
\colhead{Flux Ratio} &
\colhead{Confidence} \\
\colhead{} &
\colhead{} &
\colhead{(s)} &
\colhead{(s)} &
\colhead{(s)} &
\colhead{} &
\colhead{lower limit} &
\colhead{}
}
\startdata
  N & GRB050128 & 720.13 & 686.15 & 784.77 & 0.14 & 0.43 & 0.5163 \\
  N & GRB050128 & 293.30 & 278.18 & 305.28 & 0.09 & 0.34 & 0.4182 \\
  N & GRB050219A & 129.10 & 126.20 & 131.46 & 0.04 & 0.66 & 0.7269 \\
  N & GRB050219A & 262.85 & 245.68 & 295.79 & 0.19 & 0.72 & 0.6922 \\
  N & GRB050219A & 164.02 & 159.94 & 169.35 & 0.06 & 0.46 & 0.5689 \\
  N & GRB050318 & 32447.35 & 28612.78 & 32788.65 & 0.13 & 1.75 & 0.3661 \\
  N & GRB050319 & 1438.08 & 1376.84 & 1510.20 & 0.09 & 0.88 & 0.7775 \\
  N & GRB050401 & 139.69 & 134.39 & 151.31 & 0.12 & 0.40 & 0.5326 \\
  N & GRB050401 & 173.49 & 169.78 & 187.33 & 0.10 & 0.39 & 0.3651 \\
  N & GRB050406 & 210.50 & 112.65 & 354.36 & 1.15 & 20.42 & 0.9250 \\
  N & GRB050422 & 117.30 & 117.30 & 243.46 & 1.08 & 14.75 & 0.9557 \\
  N & GRB050502B & 749.01 & 136.66 & 1625.30 & 1.99 & 278.87 & 1.0000 \\
  N & GRB050502B & 77030.88 & 24814.44 & 148657.16 & 1.61 & 5.06 & 1.0000 \\
  N & GRB050607 & 310.98 & 278.60 & 686.64 & 1.31 & 43.21 & 1.0000 \\
  N & GRB050712 & 262.74 & 194.56 & 414.77 & 0.84 & 3.21 & 1.0000 \\
  N & GRB050712 & 478.89 & 457.87 & 546.78 & 0.19 & 4.76 & 1.0000 \\
  N & GRB050714B & 377.53 & 310.24 & 5616.99 & 14.06 & 84.09 & 1.0000 \\
  N & GRB050716 & 388.94 & 358.61 & 471.15 & 0.29 & 4.97 & 0.5335 \\
  N & GRB050716 & 174.49 & 161.11 & 194.71 & 0.19 & 0.29 & 0.4958 \\
  N & GRB050717 & 98.36 & 94.31 & 105.11 & 0.11 & 0.18 & 0.3836 \\
  N & GRB050721 & 232.62 & 219.54 & 239.21 & 0.08 & 0.26 & 0.5501 \\
  N & GRB050726 & 277.06 & 226.61 & 326.97 & 0.36 & 1.73 & 1.0000 \\
  N & GRB050726 & 163.53 & 148.62 & 177.98 & 0.18 & 0.42 & 0.5163 \\
  N & GRB050730 & 430.47 & 345.56 & 531.29 & 0.43 & 2.87 & 1.0000 \\
  N & GRB050730 & 677.47 & 615.53 & 765.81 & 0.22 & 1.23 & 1.0000 \\
  N & GRB050730 & 224.30 & 209.24 & 272.92 & 0.28 & 0.95 & 1.0000 \\
  N & GRB050803 & 726.03 & 577.64 & 908.04 & 0.46 & 0.55 & 0.8439 \\
  N & GRB050803 & 1180.30 & 1003.83 & 1247.73 & 0.21 & 0.72 & 0.7923 \\
  N & GRB050814 & 2239.74 & 1077.38 & 12353.55 & 5.03 & 2.63 & 0.7119 \\
  N & GRB050814 & 262.27 & 249.64 & 405.51 & 0.59 & 0.46 & 0.5131 \\
  N & GRB050819 & 16446.42 & 11135.25 & 36537.51 & 1.54 & 1.88 & 0.7895 \\
  N & GRB050820A & 248.09 & 215.80 & 4681.53 & 18.00 & 59.67 & 1.0000 \\
  N & GRB050822 & 424.22 & 336.28 & 945.97 & 1.44 & 40.86 & 1.0000 \\
  N & GRB050822 & 238.70 & 208.44 & 258.12 & 0.21 & 2.55 & 1.0000 \\
  N & GRB050822 & 111075.78 & 93374.52 & 150646.78 & 0.52 & 2.06 & 0.4605 \\
  N & GRB050908 & 399.39 & 294.69 & 784.93 & 1.23 & 14.23 & 1.0000 \\
  N & GRB050908 & 143.35 & 129.50 & 191.89 & 0.44 & 1.99 & 0.9211 \\
  N & GRB050915A & 105.73 & 94.78 & 156.23 & 0.58 & 4.91 & 1.0000 \\
  N & GRB050915A & 533.32 & 444.90 & 658.53 & 0.40 & 1.31 & 0.7139 \\
  N & GRB050916 & 18807.53 & 17052.58 & 22517.67 & 0.29 & 39.66 & 1.0000 \\
  N & GRB050922B & 812.98 & 615.57 & 1486.43 & 1.07 & 40.08 & 0.9196 \\
  N & GRB050922B & 375.96 & 363.91 & 391.72 & 0.07 & 0.23 & 0.7037 \\
  N & GRB051006 & 130.83 & 122.13 & 144.20 & 0.17 & 1.09 & 0.8349 \\
  N & GRB051008 & 5113.48 & 4944.48 & 5250.01 & 0.06 & 0.69 & 0.7256 \\
  N & GRB051021B & 158.89 & 126.44 & 209.28 & 0.52 & 360.69 & 0.6059 \\
  N & GRB051117A & 1324.39 & 1257.16 & 5021.08 & 2.84 & 5.60 & 1.0000 \\
  N & GRB051117A & 1072.29 & 819.59 & 1233.96 & 0.39 & 2.33 & 0.7784 \\
  N & GRB051117A & 436.58 & 301.97 & 751.52 & 1.03 & 1.80 & 0.7747 \\
  N & GRB051210 & 133.23 & 119.84 & 156.19 & 0.27 & 1.16 & 0.7904 \\
  N & GRB051210 & 164.01 & 156.19 & 217.25 & 0.37 & 0.73 & 0.4352 \\
  N & GRB051227 & 114.33 & 103.48 & 165.68 & 0.54 & 0.81 & 1.0000 \\
  N & GRB060105 & 52190.64 & 40674.94 & 98642.80 & 1.11 & 2.65 & 0.3072 \\
  N & GRB060108 & 4545.11 & 4382.39 & 4703.78 & 0.07 & 1.86 & 0.9791 \\
  N & GRB060108 & 122.79 & 122.79 & 337.44 & 1.75 & 2.86 & 0.9784 \\
  N & GRB060111A & 91.20 & 82.56 & 131.18 & 0.53 & 1.55 & 1.0000 \\
  N & GRB060111A & 168.76 & 149.80 & 204.45 & 0.32 & 2.63 & 1.0000 \\
  N & GRB060111A & 288.18 & 204.45 & 525.47 & 1.11 & 23.18 & 1.0000 \\
  N & GRB060111A & 15916.42 & 15514.31 & 16275.01 & 0.05 & 1.68 & 0.4902 \\
  N & GRB060111B & 156.34 & 136.70 & 182.76 & 0.29 & 0.48 & 0.3835 \\
  N & GRB060115 & 399.88 & 308.46 & 718.93 & 1.03 & 3.57 & 1.0000 \\
  N & GRB060116 & 179.04 & 179.04 & 209.90 & 0.17 & 1.34 & 0.9688 \\
  N & GRB060116 & 1201.25 & 1089.15 & 1356.86 & 0.22 & 0.86 & 0.7709 \\
  N & GRB060124 & 571.26 & 213.78 & 11441.52 & 19.65 & 789.65 & 1.0000 \\
  N & GRB060202 & 700.75 & 360.78 & 1040.55 & 0.97 & 5.25 & 0.9011 \\
  N & GRB060204B & 121.54 & 108.08 & 139.53 & 0.26 & 2.74 & 1.0000 \\
  N & GRB060204B & 317.43 & 275.00 & 493.00 & 0.69 & 52.47 & 1.0000 \\
  N & GRB060204B & 210.99 & 198.34 & 310.00 & 0.53 & 2.01 & 0.5519 \\
  N & GRB060206 & 5446.84 & 1787.63 & 23560.91 & 4.00 & 3.00 & 1.0000 \\
  N & GRB060210 & 199.90 & 164.69 & 302.20 & 0.69 & 12.13 & 0.8724 \\
  N & GRB060210 & 377.04 & 302.20 & 607.66 & 0.81 & 7.89 & 0.8719 \\
  N & GRB060210 & 106.86 & 104.12 & 120.23 & 0.15 & 0.86 & 0.5564 \\
  N & GRB060218 & 6473.92 & 5879.34 & 10662.93 & 0.74 & 1.30 & 0.5341 \\
  N & GRB060223A & 1319.42 & 811.26 & 5324.44 & 3.42 & 13.66 & 1.0000 \\
  N & GRB060223A & 387.14 & 292.65 & 563.22 & 0.70 & 0.97 & 0.5724 \\
Y & GRB060312 & 109.76 & 65.74 & 245.11 & 1.63 & 49.66 & 1.0000 \\
  N & GRB060312 & 542.17 & 462.80 & 850.48 & 0.72 & 1.79 & 0.9687 \\
  N & GRB060313 & 191.19 & 154.84 & 238.67 & 0.44 & 1.90 & 0.7024 \\
  N & GRB060313 & 137.14 & 120.54 & 154.84 & 0.25 & 0.79 & 0.6440 \\
  N & GRB060319 & 280.38 & 261.55 & 309.05 & 0.17 & 0.77 & 0.9771 \\
  N & GRB060403 & 73.33 & 70.07 & 79.57 & 0.13 & 0.94 & 0.5477 \\
  N & GRB060413 & 642.87 & 547.94 & 930.63 & 0.60 & 3.37 & 1.0000 \\
  N & GRB060418 & 130.77 & 116.61 & 173.07 & 0.43 & 7.22 & 0.8215 \\
  N & GRB060421 & 6045.43 & 5177.30 & 10810.05 & 0.93 & 1.28 & 0.9931 \\
  N & GRB060510A & 775.42 & 748.18 & 807.80 & 0.08 & 0.68 & 0.2755 \\
  N & GRB060510A & 1201.00 & 1171.45 & 1229.73 & 0.05 & 0.78 & 0.2447 \\
Y & GRB060510B & 301.22 & 172.14 & 468.31 & 0.98 & 13.75 & 1.0000 \\
  N & GRB060510B & 1005.71 & 751.94 & 5502.63 & 4.72 & 27.52 & 0.7393 \\
  N & GRB060512 & 201.90 & 174.23 & 379.87 & 1.02 & 3.53 & 1.0000 \\
Y & GRB060526 & 247.69 & 181.79 & 948.01 & 3.09 & 389.56 & 1.0000 \\
  N & GRB060602B & 195.41 & 174.52 & 246.99 & 0.37 & 1.19 & 0.7366 \\
  N & GRB060604 & 136.86 & 124.30 & 228.00 & 0.76 & 4.86 & 1.0000 \\
  N & GRB060607 & 98.61 & 93.41 & 132.30 & 0.39 & 4149.05 & 1.0000 \\
  N & GRB060607 & 264.86 & 216.85 & 389.35 & 0.65 & 3542.41 & 1.0000 \\
  N & GRB060607 & 180.79 & 169.89 & 205.33 & 0.20 & 1078.69 & 0.4365 \\
  N & GRB060707 & 186.08 & 175.41 & 228.78 & 0.29 & 1.41 & 0.9618 \\
  N & GRB060712 & 299.00 & 271.43 & 346.85 & 0.25 & 1.70 & 0.9314 \\
  N & GRB060714 & 137.70 & 123.58 & 158.50 & 0.25 & 3.76 & 0.8246 \\
  N & GRB060714 & 175.67 & 158.50 & 225.14 & 0.38 & 7.44 & 0.8246 \\
  N & GRB060719 & 200.98 & 139.27 & 372.09 & 1.16 & 6.68 & 1.0000 \\
  N & GRB060801 & 109.74 & 96.11 & 149.52 & 0.49 & 0.64 & 0.9223 \\
  N & GRB060805A & 4304.05 & 579.41 & 19615.79 & 4.42 & 4.97 & 0.6054 \\
  N & GRB060813 & 515.90 & 495.86 & 541.34 & 0.09 & 0.72 & 0.4417 \\
  N & GRB060813 & 109.16 & 105.35 & 129.53 & 0.22 & 0.42 & 0.4068 \\
  N & GRB060814 & 130.69 & 120.75 & 161.32 & 0.31 & 1.18 & 0.8325 \\
  N & GRB060904A & 303.95 & 253.97 & 454.43 & 0.66 & 9.92 & 1.0000 \\
  N & GRB060904A & 675.96 & 634.19 & 1036.13 & 0.59 & 6.89 & 1.0000 \\
  N & GRB060904A & 2132.63 & 1036.13 & 58529.15 & 26.96 & 8.77 & 0.9432 \\
  N & GRB060904A & 154.26 & 162.81 & 239.88 & 0.50 & -0.14 & 0.3214 \\
  N & GRB060904B & 171.72 & 127.96 & 3760.36 & 21.15 & 243.84 & 1.0000 \\
  N & GRB060906 & 162.75 & 162.75 & 244.43 & 0.50 & 4.54 & 1.0000 \\
  N & GRB060908 & 136.11 & 131.28 & 145.01 & 0.10 & 1.12 & 0.9211 \\
  N & GRB060919 & 515.99 & 353.90 & 690.68 & 0.65 & 0.91 & 0.9827 \\
  N & GRB060926 & 435.99 & 391.96 & 585.62 & 0.44 & 0.44 & 0.6280 \\
  N & GRB060929 & 553.06 & 371.65 & 1120.67 & 1.35 & 846.44 & 1.0000 \\
  N & GRB061004 & 70.54 & 70.54 & 120.35 & 0.71 & 1.15 & 0.9187 \\
  N & GRB061110A & 135.58 & 111.75 & 209.17 & 0.72 & 1.31 & 1.0000 \\
  N & GRB061121 & 80.53 & 67.01 & 100.57 & 0.42 & 0.89 & 0.9752 \\
  N & GRB061121 & 119.28 & 106.11 & 128.79 & 0.19 & 0.81 & 0.4126 \\
  N & GRB061202 & 140.58 & 125.84 & 184.32 & 0.42 & 3.74 & 1.0000 \\
  N & GRB070103 & 687.45 & 355.52 & 907.00 & 0.80 & 0.72 & 0.6708 \\
  N & GRB070107 & 357.16 & 291.36 & 399.01 & 0.30 & 9.64 & 1.0000 \\
  N & GRB070110 & 10707.35 & 4084.37 & 27535.08 & 2.19 & 8.94 & 1.0000 \\
Y & GRB070129 & 360.99 & 230.13 & 1070.24 & 2.33 & 73.58 & 1.0000 \\
  N & GRB070220 & 107.97 & 104.71 & 117.67 & 0.12 & 0.68 & 0.4355 \\
  N & GRB070220 & 523.94 & 500.25 & 580.40 & 0.15 & 0.68 & 0.3057 \\
  N & GRB070306 & 181.74 & 174.80 & 208.04 & 0.18 & 8.01 & 0.7592 \\
  N & GRB070318 & 270.70 & 235.64 & 423.34 & 0.69 & 3.06 & 1.0000 \\
  N & GRB070318 & 193.77 & 186.64 & 216.47 & 0.15 & 0.50 & 0.8203 \\
  N & GRB070330 & 222.54 & 164.56 & 361.76 & 0.89 & 11.87 & 1.0000 \\
  N & GRB070419B & 243.76 & 198.74 & 325.85 & 0.52 & 1.29 & 0.9989 \\
  N & GRB070419B & 100.23 & 86.51 & 140.86 & 0.54 & 0.31 & 0.2634 \\
  N & GRB070420 & 22932.76 & 18762.03 & 23152.81 & 0.19 & 8.38 & 1.0000 \\
  N & GRB070518 & 186.29 & 96.20 & 357.02 & 1.40 & 14.53 & 1.0000 \\
  N & GRB070520A & 235.94 & 238.83 & 3975.32 & 15.84 & -0.12 & 0.5075 \\
  N & GRB070520B & 187.56 & 146.20 & 375.30 & 1.22 & 6.66 & 1.0000 \\
  N & GRB070521 & 331.58 & 296.68 & 408.25 & 0.34 & 0.74 & 0.6976 \\
  N & GRB070531 & 427.93 & 371.50 & 558.68 & 0.44 & 0.99 & 0.7598 \\
  N & GRB070611 & 3420.85 & 3420.85 & 4131.24 & 0.21 & 1.23 & 0.9931 \\
  N & GRB070616 & 485.09 & 415.16 & 709.34 & 0.61 & 3.25 & 0.9143 \\
  N & GRB070616 & 757.27 & 713.97 & 843.16 & 0.17 & 2.03 & 0.9012 \\
  N & GRB070616 & 198.94 & 191.33 & 203.66 & 0.06 & 0.90 & 0.8182 \\
  N & GRB070621 & 145.12 & 135.99 & 154.04 & 0.12 & 0.96 & 0.5578 \\
  N & GRB070704 & 303.20 & 258.01 & 5209.41 & 16.33 & 25.52 & 1.0000 \\
  N & GRB070714A & 310.88 & 234.46 & 484.19 & 0.80 & 1.12 & 0.7697 \\
  N & GRB070714A & 904.82 & 742.63 & 1369.29 & 0.69 & 0.98 & 0.4904 \\
  N & GRB070714B & 122.80 & 114.75 & 128.57 & 0.11 & 0.85 & 0.2043 \\
  N & GRB070721B & 311.06 & 238.98 & 394.60 & 0.50 & 11.79 & 1.0000 \\
  N & GRB070721B & 623.00 & 574.66 & 748.51 & 0.28 & 0.84 & 0.7423 \\
  N & GRB070724A & 105.01 & 89.48 & 123.46 & 0.32 & 1.63 & 1.0000 \\
  N & GRB070802 & 162.48 & 162.48 & 247.07 & 0.52 & 2.47 & 0.9524 \\
  N & GRB070808 & 126.53 & 121.11 & 144.19 & 0.18 & 0.65 & 0.5954 \\
  N & GRB071031 & 454.98 & 380.95 & 6131.18 & 12.64 & 5.78 & 1.0000 \\
  N & GRB071031 & 150.28 & 141.28 & 173.11 & 0.21 & 0.52 & 0.8280 \\
  N & GRB071031 & 195.98 & 187.96 & 227.76 & 0.20 & 0.63 & 0.7122 \\
  N & GRB071031 & 257.34 & 244.61 & 300.07 & 0.22 & 1.41 & 0.6853 \\
  N & GRB071104 & 328810.48 & 32444.48 & 33294.06 & 0.00 & -0.23 & 0.5673 \\
Y & GRB071118 & 596.12 & 333.15 & 1573.09 & 2.08 & 11.63 & 0.7422 \\
  N & GRB071122 & 400.50 & 357.45 & 516.66 & 0.40 & 0.64 & 0.6866 \\
  N & GRB071227 & 158.66 & 153.12 & 191.64 & 0.24 & 0.41 & 0.7518 \\
Y & GRB080123 & 166.07 & 155.98 & 373.14 & 1.31 & 1.26 & 0.8681 \\
  N & GRB080210 & 189.07 & 174.65 & 256.70 & 0.43 & 8.61 & 1.0000 \\
Y & GRB080212 & 294.14 & 173.46 & 448.66 & 0.94 & 26.44 & 1.0000 \\
  N & GRB080229A & 104.41 & 93.74 & 173.11 & 0.76 & 0.98 & 0.9438 \\
Y & GRB080310 & 205.60 & 126.20 & 1131.56 & 4.89 & 41.66 & 1.0000 \\
  N & GRB080310 & 4858.46 & 1442.36 & 17540.08 & 3.31 & 3.18 & 0.5355 \\
  N & GRB080319B & 701334.41 & 76075.39 & 2544607.50 & 3.52 & 12.28 & 0.5956 \\
  N & GRB080319D & 295.75 & 238.44 & 490.77 & 0.85 & 6.46 & 0.9597 \\
  N & GRB080320 & 309.90 & 273.92 & 444.89 & 0.55 & 6.68 & 1.0000 \\
  N & GRB080320 & 211.29 & 196.35 & 273.92 & 0.37 & 3.21 & 0.6243 \\
  N & GRB080320 & 699.37 & 759.78 & 972.75 & 0.30 & -0.33 & 0.6198 \\
  N & GRB080325 & 220.02 & 198.71 & 379.57 & 0.82 & 2.01 & 1.0000 \\
  N & GRB080325 & 175.59 & 163.52 & 175.59 & 0.07 & -0.10 & 0.8691 \\
  N & GRB080409 & 441.94 & 342.75 & 5615.39 & 11.93 & 0.92 & 0.7722 \\
  N & GRB080426 & 653.06 & 572.11 & 736.85 & 0.25 & 0.72 & 0.9387 \\
Y & GRB080506 & 480.47 & 366.33 & 5652.40 & 11.00 & 43.69 & 1.0000 \\
  N & GRB080506 & 174.65 & 160.37 & 207.62 & 0.27 & 1.07 & 0.7826 \\
  N & GRB080516 & 463.15 & 356.73 & 572.61 & 0.47 & 0.89 & 0.7085 \\
  N & GRB080517 & 131.45 & 131.45 & 565.45 & 3.30 & 18.58 & 0.9385 \\
  N & GRB080602 & 906.16 & 890.47 & 922.79 & 0.04 & 1.53 & 0.6063 \\
  N & GRB080604 & 972.20 & 896.44 & 1065.04 & 0.17 & 0.85 & 0.3884 \\
  N & GRB080607 & 123.97 & 117.35 & 209.98 & 0.75 & 5.45 & 1.0000 \\
  N & GRB080703 & 380.77 & 306.95 & 420.32 & 0.30 & 0.66 & 0.6119 \\
  N & GRB080703 & 217.34 & 189.30 & 306.95 & 0.54 & 0.53 & 0.4867 \\
  N & GRB080710 & 3467.74 & 3314.73 & 4716.99 & 0.40 & 0.58 & 0.5107 \\
  N & GRB080714 & 165.06 & 143.53 & 170.17 & 0.16 & 1.30 & 0.4506 \\
  N & GRB080723A & 160.30 & 133.13 & 193.33 & 0.38 & 0.69 & 0.9053 \\
  N & GRB080727A & 317.82 & 243.40 & 426.37 & 0.58 & 0.93 & 0.9880 \\
  N & GRB080802 & 94.21 & 86.43 & 104.05 & 0.19 & 2.80 & 1.0000 \\
  N & GRB080804 & 117.49 & 114.24 & 121.16 & 0.06 & 0.50 & 0.7253 \\
  N & GRB080804 & 137.47 & 133.67 & 145.29 & 0.08 & 0.33 & 0.5335 \\
Y & GRB080805 & 120.25 & 89.90 & 224.96 & 1.12 & 4.70 & 1.0000 \\
  N & GRB080810 & 103.46 & 88.29 & 132.09 & 0.42 & 4.54 & 1.0000 \\
Y & GRB080810 & 208.00 & 187.38 & 335.21 & 0.71 & 5.38 & 1.0000 \\
  N & GRB080905A & 256.22 & 200.92 & 310.51 & 0.43 & 0.81 & 0.7808 \\
  N & GRB080906 & 180.59 & 160.85 & 257.76 & 0.54 & 2.17 & 1.0000 \\
  N & GRB080906 & 577.76 & 552.98 & 709.40 & 0.27 & 1.07 & 0.8932 \\
  N & GRB080913 & 1863.81 & 984.59 & 8580.70 & 4.08 & 6.46 & 0.9997 \\
  N & GRB080913 & 485.17 & 305.21 & 608.29 & 0.62 & 1.24 & 0.7351 \\
  N & GRB080916A & 92.22 & 91.42 & 97.82 & 0.07 & 0.74 & 0.7920 \\
  N & GRB080919 & 285.07 & 206.63 & 711.80 & 1.77 & 5.82 & 0.9984 \\
  N & GRB080928 & 206.76 & 176.27 & 269.13 & 0.45 & 3.94 & 1.0000 \\
  N & GRB080928 & 349.63 & 343.42 & 400.61 & 0.16 & 2.07 & 1.0000 \\
  N & GRB081008 & 301.44 & 284.17 & 403.08 & 0.39 & 5.90 & 1.0000 \\
Y & GRB081008 & 171.82 & 126.22 & 237.45 & 0.65 & 2.05 & 0.7415 \\
  N & GRB081011 & 113.45 & 113.45 & 208.35 & 0.84 & 5.85 & 1.0000 \\
  N & GRB081011 & 52704.36 & 44311.60 & 72949.15 & 0.54 & 1.03 & 0.6849 \\
  N & GRB081024A & 168.52 & 129.46 & 254.25 & 0.74 & 13.52 & 0.9330 \\
  N & GRB081102 & 954.84 & 876.46 & 5796.52 & 5.15 & 30.54 & 1.0000 \\
  N & GRB081121 & 3645.38 & 3442.15 & 3913.02 & 0.13 & 0.97 & 0.5174 \\
  N & GRB081128 & 32368.08 & 14915.69 & 72762.53 & 1.79 & 1.80 & 0.4221 \\
Y & GRB081210 & 141.11 & 111.54 & 249.65 & 0.98 & 8.48 & 1.0000 \\
  N & GRB081210 & 316.95 & 282.45 & 471.49 & 0.60 & 1.78 & 0.8943 \\
  N & GRB090111 & 474.40 & 174.38 & 1201.42 & 2.16 & 20.85 & 1.0000 \\
  N & GRB090123 & 1781.95 & 1452.29 & 1912.69 & 0.26 & 1.15 & 0.7904 \\
  N & GRB090123 & 483.95 & 416.29 & 667.62 & 0.52 & 1.35 & 0.4320 \\
  N & GRB090309 & 4741.20 & 4352.06 & 16512.02 & 2.56 & 0.89 & 0.6757 \\
  N & GRB090328A & 93405.55 & 63169.38 & 98447.09 & 0.38 & 1.24 & 0.5401 \\
Y & GRB090407 & 134.67 & 111.10 & 746.38 & 4.72 & 12.41 & 1.0000 \\
  N & GRB090417B & 1507.40 & 1244.48 & 5401.39 & 2.76 & 11.91 & 1.0000 \\
  N & GRB090418A & 157.92 & 152.74 & 172.02 & 0.12 & 0.31 & 0.3574 \\
Y & GRB090419 & 321.96 & 265.96 & 1015.39 & 2.33 & 4.18 & 0.9837 \\
  N & GRB090422 & 96.73 & 73.85 & 120.62 & 0.48 & 0.82 & 0.9881 \\
  N & GRB090423 & 174.42 & 136.90 & 273.99 & 0.79 & 13.22 & 1.0000 \\
  N & GRB090426 & 296.20 & 200.23 & 417.24 & 0.73 & 0.60 & 0.6446 \\
  N & GRB090429A & 169.05 & 156.27 & 209.16 & 0.31 & 2.77 & 0.9287 \\
  N & GRB090429A & 101.24 & 88.71 & 117.80 & 0.29 & 0.84 & 0.7504 \\
  N & GRB090429A & 130.17 & 120.73 & 138.18 & 0.13 & 0.75 & 0.6688 \\
  N & GRB090429A & 252.31 & 241.35 & 37669.76 & 148.34 & 1.23 & 0.5717 \\
  N & GRB090429B & 626.04 & 290.50 & 15542.75 & 24.36 & 3.86 & 0.9746 \\
  N & GRB090515 & 159.26 & 79.51 & 288.97 & 1.32 & 16.00 & 1.0000 \\
  N & GRB090515 & 204.58 & 201.16 & 207.78 & 0.03 & 11386341.84 & 0.2487 \\
  N & GRB090516 & 274.00 & 267.27 & 282.37 & 0.06 & 5.95 & 1.0000 \\
  N & GRB090519 & 221.48 & 203.87 & 242.48 & 0.17 & 1.27 & 0.8710 \\
  N & GRB090529 & 13785.24 & 27785.00 & 148937.11 & 8.79 & 0.21 & 0.8682 \\
  N & GRB090530 & 263.97 & 192.74 & 303.82 & 0.42 & 0.71 & 0.8561 \\
  N & GRB090607 & 119.71 & 96.30 & 549.87 & 3.79 & 7.91 & 1.0000 \\
  N & GRB090621A & 268.16 & 153.69 & 795.78 & 2.39 & 552.54 & 1.0000 \\
  N & GRB090628 & 14599.13 & 3522.28 & 42257.78 & 2.65 & 4.21 & 0.7659 \\
  N & GRB090709A & 88.71 & 79.38 & 105.82 & 0.30 & 3.10 & 1.0000 \\
  N & GRB090709A & 399.78 & 366.96 & 466.43 & 0.25 & 0.91 & 0.8495 \\
  N & GRB090709A & 281.83 & 248.49 & 307.03 & 0.21 & 0.65 & 0.6830 \\
Y & GRB090715B & 289.92 & 59.33 & 363.63 & 1.05 & 30.31 & 1.0000 \\
  N & GRB090727 & 269.62 & 131.63 & 4078.65 & 14.64 & 109.21 & 1.0000 \\
  N & GRB090727 & 171920.17 & 113699.93 & 285295.30 & 1.00 & 1.43 & 0.6279 \\
  N & GRB090728 & 227.44 & 194.06 & 315.61 & 0.53 & 0.80 & 0.4978 \\
  N & GRB090809 & 178.61 & 170.45 & 3512.68 & 18.71 & 15.04 & 1.0000 \\
  N & GRB090809 & 4702.60 & 3628.26 & 9240.61 & 1.19 & 8.73 & 1.0000 \\
  N & GRB090809 & 22799.20 & 11459.20 & 31007.17 & 0.86 & 1.05 & 0.6837 \\
  N & GRB090812 & 137.77 & 102.78 & 199.20 & 0.70 & 0.66 & 0.9197 \\
  N & GRB090812 & 257.63 & 246.38 & 318.82 & 0.28 & 1.78 & 0.8979 \\
  N & GRB090831C & 182.28 & 155.56 & 296.96 & 0.78 & 7.05 & 0.9134 \\
  N & GRB090831C & 431.80 & 371.25 & 617.50 & 0.57 & 2.38 & 0.7988 \\
  N & GRB090902B & 436259.21 & 338770.66 & 703452.43 & 0.84 & 1.32 & 0.5544 \\
  N & GRB090904A & 300.01 & 288.54 & 358.19 & 0.23 & 2.76 & 0.9685 \\
  N & GRB090904A & 6758.74 & 6107.89 & 10532.16 & 0.65 & 0.93 & 0.3939 \\
  N & GRB090904B & 120.51 & 120.51 & 145.20 & 0.20 & 3.38 & 0.9745 \\
  N & GRB090904B & 903.80 & 803.70 & 967.61 & 0.18 & 1.00 & 0.5399 \\
  N & GRB090912 & 781.36 & 781.36 & 850.68 & 0.09 & 2.26 & 0.8477 \\
  N & GRB090926A & 46846.73 & 46677.98 & 51695.21 & 0.11 & 0.78 & 0.5713 \\
  N & GRB090926A & 203029.11 & 193054.42 & 232375.73 & 0.19 & 0.89 & 0.3281 \\
  N & GRB090926A & 86116.04 & 81266.45 & 144410.24 & 0.73 & 1.71 & 0.2109 \\
  N & GRB090927 & 2229.09 & 2229.09 & 2649.06 & 0.19 & 0.87 & 0.9198 \\
Y & GRB090929B & 149.58 & 81.59 & 3453.39 & 22.54 & 13.26 & 1.0000 \\
  N & GRB091024 & 5145.00 & 4606.70 & 32019.65 & 5.33 & 1.63 & 0.9204 \\
  N & GRB091024 & 3207.78 & 3207.78 & 3395.99 & 0.06 & 1.92 & 0.7846 \\
  N & GRB091026 & 341.14 & 288.89 & 545.75 & 0.75 & 6.52 & 0.9932 \\
  N & GRB091026 & 877.11 & 618.25 & 5557.31 & 5.63 & 3.06 & 0.7291 \\
  N & GRB091026 & 173.04 & 158.18 & 222.06 & 0.37 & 1.20 & 0.5386 \\
  N & GRB091029 & 323.60 & 233.28 & 601.56 & 1.14 & 8.67 & 0.8354 \\
  N & GRB091104 & 203.85 & 191.11 & 250.03 & 0.29 & 1.52 & 0.8765 \\
  N & GRB091109A & 248.34 & 227.27 & 278.06 & 0.20 & 0.71 & 0.6050 \\
  N & GRB091130B & 99.92 & 83.85 & 171.67 & 0.88 & 3.15 & 0.9167 \\
  N & GRB091208B & 101.40 & 101.40 & 144.38 & 0.42 & 0.93 & 0.8409 \\
  N & GRB091221 & 106.33 & 87.64 & 196.18 & 1.02 & 10.91 & 0.7838 \\
  N & GRB091221 & 62.99 & 62.99 & 87.64 & 0.39 & 5.96 & 0.7721 \\
  N & GRB100111A & 951.32 & 817.28 & 4644.77 & 4.02 & 0.80 & 0.9311 \\
  N & GRB100117A & 181.60 & 150.75 & 470.16 & 1.76 & 2.35 & 0.7065 \\
Y & GRB100212A & 120.43 & 64.70 & 459.25 & 3.28 & 21.32 & 1.0000 \\
  N & GRB100212A & 668.35 & 610.33 & 795.63 & 0.28 & 12.29 & 0.7407 \\
  N & GRB100219A & 17800.44 & 13234.12 & 18999.62 & 0.32 & 0.98 & 0.7968 \\
Y & GRB100302A & 250.31 & 225.14 & 780.53 & 2.22 & 13.32 & 1.0000 \\
  N & GRB100302A & 134.48 & 123.44 & 150.19 & 0.20 & 1.51 & 0.8935 \\
  N & GRB100302A & 188.31 & 174.49 & 201.33 & 0.14 & 1.40 & 0.7201 \\
  N & GRB100316B & 1064.41 & 395.16 & 37185.71 & 34.56 & 2.82 & 0.9665 \\
  N & GRB100316C & 261.01 & 188.34 & 574.54 & 1.48 & 1.14 & 0.9899 \\
  N & GRB100413A & 149.91 & 138.19 & 160.53 & 0.15 & 0.90 & 0.7837 \\
  N & GRB100413A & 278.44 & 250.86 & 289.21 & 0.14 & 1.05 & 0.7793 \\
  N & GRB100413A & 222.48 & 192.98 & 234.85 & 0.19 & 0.83 & 0.5932 \\
  N & GRB100425A & 70.97 & 69.85 & 85.61 & 0.22 & 2.54 & 0.8257 \\
  N & GRB100425A & 482.86 & 347.57 & 610.09 & 0.54 & 0.95 & 0.5149 \\
  N & GRB100504A & 52.96 & 52.96 & 69.43 & 0.31 & 10.39 & 1.0000 \\
  N & GRB100504A & 81.58 & 74.95 & 117.28 & 0.52 & 1.02 & 0.8652 \\
  N & GRB100513A & 213.64 & 163.43 & 698.32 & 2.50 & 5.83 & 0.9078 \\
  N & GRB100522A & 2021.97 & 1703.73 & 5996.03 & 2.12 & 0.46 & 0.4785 \\
  N & GRB100526A & 183.95 & 168.61 & 259.30 & 0.49 & 1.80 & 0.9893 \\
  N & GRB100614A & 161.92 & 154.70 & 180.64 & 0.16 & 0.63 & 0.9164 \\
  N & GRB100614A & 957.82 & 898.59 & 1163.40 & 0.28 & 0.76 & 0.5811 \\
  N & GRB100619A & 941.53 & 862.01 & 5001.70 & 4.40 & 71.41 & 1.0000 \\
  N & GRB100619A & 88.37 & 72.86 & 123.36 & 0.57 & 6.76 & 0.9822 \\
  N & GRB100621A & 65.92 & 65.92 & 69.85 & 0.06 & 3.53 & 0.9805 \\
  N & GRB100625A & 191.48 & 134.47 & 326.50 & 1.00 & 0.69 & 0.5371 \\
  N & GRB100702A & 361.05 & 237.42 & 442.05 & 0.57 & 1.19 & 0.3849 \\
Y & GRB100704A & 173.54 & 145.30 & 348.81 & 1.17 & 16.89 & 0.9734 \\
  N & GRB100725A & 69.18 & 69.18 & 73.55 & 0.06 & 2.04 & 0.7670 \\
Y & GRB100725B & 217.54 & 114.22 & 369.85 & 1.18 & 56.43 & 0.7892 \\
  N & GRB100727A & 243.37 & 163.06 & 669.58 & 2.08 & 63.09 & 1.0000 \\
  N & GRB100728A & 574.10 & 512.64 & 654.75 & 0.25 & 6.75 & 0.8812 \\
  N & GRB100728A & 317.33 & 298.03 & 380.33 & 0.26 & 4.43 & 0.8121 \\
  N & GRB100728A & 701.40 & 673.31 & 886.50 & 0.30 & 2.50 & 0.7756 \\
  N & GRB100728A & 123.41 & 109.40 & 137.56 & 0.23 & 0.93 & 0.7111 \\
  N & GRB100728A & 392.83 & 380.33 & 415.01 & 0.09 & 3.55 & 0.5152 \\
  N & GRB100728A & 221.84 & 197.98 & 247.34 & 0.22 & 1.01 & 0.5001 \\
  N & GRB100728A & 88.39 & 82.89 & 100.06 & 0.19 & 0.48 & 0.3698 \\
  N & GRB100728A & 462.31 & 448.44 & 498.57 & 0.11 & 1.14 & 0.2746 \\
  N & GRB100728A & 269.42 & 251.77 & 292.42 & 0.15 & 0.88 & 0.2122 \\
  N & GRB100728B & 104.13 & 88.37 & 144.33 & 0.54 & 0.65 & 0.6330 \\
  N & GRB100802A & 478.02 & 274.48 & 4633.13 & 9.12 & 47.29 & 1.0000 \\
  N & GRB100802A & 33392.37 & 29132.41 & 40373.93 & 0.34 & 1.55 & 0.4487 \\
  N & GRB100805A & 636.29 & 423.70 & 4348.88 & 6.17 & 23.06 & 1.0000 \\
  N & GRB100807A & 88.35 & 77.18 & 262.37 & 2.10 & 27.40 & 1.0000 \\
  N & GRB100814A & 147.00 & 119.46 & 357.38 & 1.62 & 2.18 & 0.9918 \\
  N & GRB100814A & 139311.60 & 69791.07 & 374420.00 & 2.19 & 1.81 & 0.6527 \\
  N & GRB100816A & 73.82 & 73.82 & 91.72 & 0.24 & 1.05 & 0.8600 \\
  N & GRB100816A & 139.65 & 125.09 & 211.06 & 0.62 & 1.24 & 0.8562 \\
  N & GRB100823A & 4748.51 & 4398.85 & 5317.38 & 0.19 & 2.07 & 0.4737 \\
Y & GRB100901A & 399.18 & 132.52 & 3851.99 & 9.32 & 42.59 & 1.0000 \\
  N & GRB100901A & 28505.81 & 12080.08 & 67586.83 & 1.95 & 1.68 & 0.8896 \\
  N & GRB100902A & 411.02 & 355.83 & 634.69 & 0.68 & 131.92 & 1.0000 \\
  N & GRB100902A & 2015.83 & 1846.22 & 2194.78 & 0.17 & 1.09 & 0.5733 \\
Y & GRB100905A & 319.46 & 161.44 & 563.72 & 1.26 & 47.64 & 1.0000 \\
  N & GRB100905A & 5412.16 & 1850.27 & 7200.81 & 0.99 & 1.13 & 0.5858 \\
  N & GRB100905A & 1683.70 & 1535.22 & 1850.70 & 0.19 & 0.95 & 0.4221 \\
  N & GRB100906A & 117.90 & 86.15 & 199.41 & 0.96 & 23.16 & 1.0000 \\
  N & GRB100915A & 157.84 & 153.95 & 166.30 & 0.08 & 0.34 & 0.6480 \\
  N & GRB100915A & 191.01 & 179.20 & 194.34 & 0.08 & 0.78 & 0.3489 \\
  N & GRB101011A & 120.10 & 108.08 & 143.44 & 0.29 & 0.65 & 0.6533 \\
  N & GRB101011A & 241.37 & 215.77 & 296.95 & 0.34 & 0.98 & 0.4961 \\
  N & GRB101017A & 849.85 & 679.06 & 1058.20 & 0.45 & 0.60 & 0.5488 \\
  N & GRB101017A & 181.22 & 173.93 & 193.02 & 0.11 & 0.71 & 0.3320 \\
  N & GRB101023A & 75.25 & 75.25 & 77.63 & 0.03 & 8.14 & 0.9456 \\
  N & GRB101024A & 513.90 & 482.84 & 576.60 & 0.18 & 0.33 & 0.3508 \\
  N & GRB101030A & 86.68 & 84.33 & 89.96 & 0.06 & 0.26 & 0.4242 \\
  N & GRB101117B & 184.22 & 161.53 & 244.90 & 0.45 & 0.44 & 0.6498 \\
  N & GRB101117B & 266.06 & 244.90 & 312.12 & 0.25 & 0.80 & 0.5989 \\
  N & GRB101204A & 316164.03 & 290600.72 & 334743.34 & 0.14 & 0.88 & 0.3346 \\
  N & GRB101213A & 95.34 & 95.34 & 98.27 & 0.03 & 0.45 & 0.5303 \\
  N & GRB101213A & 63343.55 & 34883.14 & 80229.22 & 0.72 & 1.36 & 0.5222 \\
  N & GRB101219A & 170.16 & 84.25 & 270.53 & 1.09 & 7.92 & 0.9501 \\
  N & GRB101219B & 327.66 & 300.05 & 554.06 & 0.78 & 0.94 & 0.6871 \\
  N & GRB101225A & 22590.84 & 10740.74 & 56608.60 & 2.03 & 34.28 & 0.8784 \\
  N & GRB101225A & 6108.55 & 4988.46 & 7477.77 & 0.41 & 5.03 & 0.7767 \\
Y & GRB110102A & 263.18 & 193.02 & 443.52 & 0.95 & 38.17 & 1.0000 \\
  N & GRB110102A & 139.57 & 139.57 & 178.35 & 0.28 & 16.64 & 1.0000 \\
  N & GRB110106B & 829.59 & 583.60 & 1111.41 & 0.64 & 0.62 & 0.7593 \\
  N & GRB110112A & 752.34 & 601.82 & 4917.12 & 5.74 & 0.89 & 0.9922 \\
  N & GRB110112A & 268.00 & 186.65 & 447.82 & 0.97 & 0.65 & 0.8104 \\
  N & GRB110119A & 197.09 & 164.64 & 305.24 & 0.71 & 0.11 & 0.9121 \\
  N & GRB110119A & 48.53 & 48.53 & 68.56 & 0.41 & 1.69 & 0.8997 \\
  N & GRB110119A & 127.42 & 99.59 & 147.52 & 0.38 & 2.03 & 0.8311 \\
  N & GRB110119A & 657.91 & 533.00 & 4942.73 & 6.70 & 6.81 & 0.7989 \\
  N & GRB110119A & 385.41 & 364.54 & 441.68 & 0.20 & 2.36 & 0.5654 \\
  N & GRB110128A & 129.73 & 129.73 & 315.19 & 1.43 & 2.02 & 0.9991 \\
  N & GRB110128A & 171625.12 & 83377.92 & 217445.33 & 0.78 & 1.12 & 0.5554 \\
  N & GRB110201A & 148.34 & 110.01 & 308.08 & 1.34 & 2.19 & 0.7542 \\
  N & GRB110205A & 615.56 & 595.33 & 716.93 & 0.20 & 1.99 & 0.8407 \\
  N & GRB110205A & 82840.56 & 76124.31 & 118778.54 & 0.51 & 0.81 & 0.2399 \\
  N & GRB110208A & 68.62 & 68.62 & 143.90 & 1.10 & 2.96 & 0.9999 \\
  N & GRB110208A & 844.81 & 617.34 & 5677.35 & 5.99 & 0.71 & 0.7111 \\
  N & GRB110213A & 98.78 & 87.04 & 103.59 & 0.17 & 0.75 & 0.4149 \\
  N & GRB110223A & 275.06 & 188.23 & 353.46 & 0.60 & 0.80 & 0.7730 \\
  N & GRB110223B & 65.15 & 58.32 & 77.63 & 0.30 & 1.72 & 0.8741 \\
  N & GRB110223B & 1173.23 & 1075.00 & 1235.58 & 0.14 & 1.18 & 0.6455 \\
  N & GRB110305A & 358.97 & 253.95 & 638.55 & 1.07 & 1.05 & 0.5774 \\
  N & GRB110312A & 154.00 & 148.94 & 168.46 & 0.13 & 1.13 & 0.9126 \\
  N & GRB110312A & 423.56 & 322.33 & 529.20 & 0.49 & 0.77 & 0.7002 \\
  N & GRB110312A & 737.88 & 646.25 & 808.11 & 0.22 & 0.53 & 0.3664 \\
  N & GRB110312A & 216773.86 & 141370.86 & 363358.27 & 1.02 & 0.92 & 0.2426 \\
  N & GRB110315A & 514.75 & 411.47 & 3796.30 & 6.58 & 8.86 & 0.9969 \\
  N & GRB110318B & 140.05 & 110.67 & 275.63 & 1.18 & 1.98 & 0.9708 \\
  N & GRB110319A & 65.96 & 62.89 & 67.85 & 0.08 & 0.37 & 0.3692 \\
  N & GRB110407A & 436.34 & 401.55 & 802.43 & 0.92 & -0.29 & 1.0000 \\
  N & GRB110407A & 4967.67 & 4448.41 & 5294.80 & 0.17 & -0.66 & 0.4489 \\
  N & GRB110414A & 385.10 & 275.54 & 648.81 & 0.97 & 7.05 & 0.8878 \\
  N & GRB110414A & 155.69 & 139.68 & 172.83 & 0.21 & 0.80 & 0.6283 \\
  N & GRB110520A & 258.16 & 146.39 & 494.65 & 1.35 & 6.61 & 0.9241 \\
  N & GRB110520A & 625.88 & 494.65 & 737.50 & 0.39 & 0.96 & 0.7595 \\
  N & GRB110521A & 184.23 & 215.74 & 524.90 & 1.68 & 0.43 & 0.7281 \\
  N & GRB110530A & 7344.57 & 2560.24 & 13542.38 & 1.50 & 0.57 & 0.7548 \\
  N & GRB110530A & 1341.81 & 1009.04 & 1718.77 & 0.53 & 1.25 & 0.6712 \\
  N & GRB110610A & 217.98 & 179.12 & 291.11 & 0.51 & 1.11 & 0.7112 \\
  N & GRB110610A & 653.14 & 614.80 & 811.51 & 0.30 & 1.83 & 0.6782 \\
  N & GRB110610A & 199925.99 & 119058.16 & 298760.95 & 0.90 & 1.34 & 0.6189 \\
  N & GRB110625A & 726.77 & 685.98 & 829.53 & 0.20 & 0.77 & 0.4342 \\
  N & GRB110709A & 56.12 & 56.12 & 75.16 & 0.34 & 1.66 & 1.0000 \\
  N & GRB110709A & 91.55 & 87.76 & 109.09 & 0.23 & 0.96 & 0.6242 \\
Y & GRB110709B & 650.15 & 417.88 & 1522.12 & 1.70 & 37.91 & 1.0000 \\
  N & GRB110709B & 70.72 & 70.72 & 112.03 & 0.58 & 4.06 & 0.7481 \\
  N & GRB110709B & 157.67 & 133.39 & 256.11 & 0.78 & 0.76 & 0.7203 \\
  N & GRB110715A & 50165.92 & 35049.31 & 257277.36 & 4.43 & 2.92 & 0.5119 \\
  N & GRB110726A & 398.09 & 354.92 & 492.30 & 0.35 & 1.16 & 0.9631 \\
  N & GRB110726A & 52.74 & 48.97 & 117.52 & 1.30 & 0.42 & 0.9532 \\
  N & GRB110801A & 382.04 & 322.01 & 774.21 & 1.18 & 77.04 & 1.0000 \\
  N & GRB110801A & 213.18 & 176.15 & 261.79 & 0.40 & 2.21 & 0.6649 \\
  N & GRB110820A & 269.30 & 142.11 & 558.35 & 1.55 & 1072.55 & 1.0000 \\
  N & GRB110915A & 161.67 & 148.71 & 216.96 & 0.42 & 1.41 & 0.7723 \\
  N & GRB110921A & 224.20 & 155.43 & 403.30 & 1.11 & 3.71 & 0.7655 \\
  N & GRB110921A & 526.15 & 403.30 & 830.05 & 0.81 & 5.92 & 0.6207 \\
  N & GRB110921A & 1285.70 & 1182.23 & 1346.57 & 0.13 & 0.93 & 0.3762 \\
Y & GRB111016A & 610.62 & 387.58 & 4950.64 & 7.47 & 119.78 & 1.0000 \\
  N & GRB111018A & 118.30 & 118.30 & 191.88 & 0.62 & 7.10 & 0.9997 \\
  N & GRB111020A & 904.74 & 557.83 & 1204.88 & 0.72 & 1.05 & 0.6795 \\
  N & GRB111020A & 34831.61 & 25977.80 & 52426.19 & 0.76 & 1.22 & 0.5243 \\
  N & GRB111022B & 455.80 & 410.88 & 521.26 & 0.24 & 0.63 & 0.5284 \\
Y & GRB111103B & 115.88 & 110.88 & 4672.68 & 39.37 & 30.10 & 1.0000 \\
  N & GRB111107A & 332.37 & 91.67 & 587.32 & 1.49 & 6.95 & 0.6913 \\
  N & GRB111117A & 150.16 & 107.95 & 247.29 & 0.93 & 0.59 & 0.7561 \\
  N & GRB111123A & 487.90 & 466.14 & 656.27 & 0.39 & 1.91 & 0.9799 \\
  N & GRB111123A & 285.37 & 267.58 & 300.85 & 0.12 & 0.50 & 0.5189 \\
  N & GRB111123A & 146.21 & 139.44 & 157.62 & 0.12 & 0.34 & 0.4287 \\
  N & GRB111129A & 254.03 & 194.19 & 357.28 & 0.64 & 0.56 & 0.6207 \\
  N & GRB120102A & 1074.69 & 937.13 & 10435.95 & 8.84 & 13.79 & 0.9622 \\
  N & GRB120121A & 108.03 & 357.28 & 154.60 & -1.88 & 4.91 & 0.9966 \\
  N & GRB120121A & 1451.90 & 928.72 & 6499.65 & 3.84 & 0.96 & 0.5121 \\
  N & GRB120213A & 5511.00 & 957.62 & 11236.18 & 1.87 & 2.52 & 0.8825 \\
  N & GRB120224A & 99.14 & 99.14 & 194.32 & 0.96 & 30.91 & 0.9996 \\
  N & GRB120224A & 1116.54 & 582.13 & 5248.08 & 4.18 & 1.94 & 0.8922 \\
  N & GRB120305A & 117.96 & 86.36 & 310.19 & 1.90 & 2.15 & 0.9957 \\
Y & GRB120308A & 123.68 & 111.40 & 316.02 & 1.65 & 6.91 & 1.0000 \\
  N & GRB120308A & 2386.69 & 1932.85 & 5891.78 & 1.66 & 0.63 & 0.4988 \\
  N & GRB120311A & 66716.90 & 33092.84 & 104785.26 & 1.07 & 0.89 & 0.9656 \\
  N & GRB120312A & 90.91 & 90.91 & 132.31 & 0.46 & 2.01 & 0.9973 \\
  N & GRB120320A & 161.22 & 161.22 & 312.33 & 0.94 & 52.97 & 0.9995 \\
  N & GRB120320A & 41696.11 & 16636.81 & 139911.52 & 2.96 & 3.10 & 0.5723 \\
  N & GRB120324A & 102.84 & 99.97 & 120.89 & 0.20 & 1.09 & 0.8069 \\
  N & GRB120324A & 40264.89 & 17496.23 & 68368.09 & 1.26 & -0.24 & 0.7154 \\
  N & GRB120327A & 931.16 & 856.83 & 963.57 & 0.11 & 0.79 & 0.3666 \\
  N & GRB120328A & 63.45 & 63.45 & 88.04 & 0.39 & 3.62 & 0.9258 \\
  N & GRB120328A & 123.19 & 88.04 & 232.50 & 1.17 & 9.99 & 0.9121 \\
  N & GRB120328A & 551.99 & 366.34 & 783.08 & 0.75 & 1.67 & 0.6212 \\
  N & GRB120401A & 239.16 & 196.14 & 350.84 & 0.65 & 0.82 & 0.6788 \\
  N & GRB120401A & 105.55 & 105.55 & 107.95 & 0.02 & 0.78 & 0.4007 \\
  N & GRB120514A & 142.80 & 126.64 & 471.16 & 2.41 & 13.47 & 0.9998 \\
  N & GRB120514A & 643.81 & 471.16 & 4497.94 & 6.25 & 7.07 & 0.6668 \\
  N & GRB120521A & 216.64 & 98.67 & 325.92 & 1.05 & 1.77 & 0.9789 \\
  N & GRB120521B & 6380.21 & 5604.57 & 7181.97 & 0.25 & 1.81 & 0.3365 \\
  N & GRB120612A & 4956.31 & 1337.61 & 17301.26 & 3.22 & 8.38 & 0.9224 \\
  N & GRB120612A & 181.49 & 171.17 & 193.16 & 0.12 & 1.57 & 0.5148 \\
  N & GRB120701A & 305.16 & 273.31 & 459.20 & 0.61 & 1.21 & 0.8231 \\
  N & GRB120701A & 634.48 & 538.54 & 6076.92 & 8.73 & 0.83 & 0.5857 \\
  N & GRB120703A & 72.54 & 72.54 & 92.46 & 0.27 & 1.70 & 0.9290 \\
  N & GRB120703A & 177.49 & 159.99 & 193.41 & 0.19 & 0.84 & 0.6880 \\
  N & GRB120703A & 232.24 & 214.10 & 252.48 & 0.17 & 0.84 & 0.3913 \\
  N & GRB120711B & 805.15 & 443.94 & 892.78 & 0.56 & 2.78 & 0.9424 \\
  N & GRB120711B & 357.80 & 341.53 & 417.47 & 0.21 & 0.93 & 0.6084 \\
  N & GRB120712A & 126970.53 & 67633.74 & 462512.44 & 3.11 & 2.60 & 0.7872 \\
  N & GRB120714A & 29832.97 & 21938.09 & 42777.41 & 0.70 & 0.59 & 0.3589 \\
  N & GRB120722A & 300.58 & 217.42 & 558.65 & 1.14 & 3.59 & 0.8054 \\
  N & GRB120722A & 141.60 & 141.60 & 217.42 & 0.54 & 6.66 & 0.7907 \\
  N & GRB120724A & 119.95 & 117.16 & 129.34 & 0.10 & 0.43 & 0.5839 \\
  N & GRB120728A & 136.62 & 136.62 & 236.80 & 0.73 & 15.00 & 0.9962 \\
  N & GRB120728A & 548.01 & 357.73 & 1061.73 & 1.28 & 18.40 & 0.8091 \\
  N & GRB120729A & 513.22 & 442.45 & 565.39 & 0.24 & 0.52 & 0.5557 \\
  N & GRB120729A & 95.58 & 90.39 & 97.60 & 0.08 & 0.68 & 0.5227 \\
  N & GRB120804A & 82.82 & 82.82 & 104.29 & 0.26 & 1.13 & 0.9142 \\
  N & GRB120804A & 289.82 & 260.75 & 418.23 & 0.54 & 0.86 & 0.6700 \\
  N & GRB120807A & 115.47 & 101.13 & 221.15 & 1.04 & 6.93 & 0.9920 \\
  N & GRB120816A & 210.72 & 172.29 & 257.86 & 0.41 & 0.89 & 0.6112 \\
  N & GRB120816A & 497.47 & 383.06 & 580.38 & 0.40 & 0.87 & 0.3832 \\
  N & GRB120907A & 175.19 & 121.88 & 311.06 & 1.08 & 0.48 & 0.6776 \\
  N & GRB120911A & 4752.70 & 4548.38 & 5063.46 & 0.11 & 1.10 & 0.7967 \\
  N & GRB120922A & 324.66 & 305.30 & 349.92 & 0.14 & 1.56 & 0.5833 \\
  N & GRB120922A & 411.67 & 361.28 & 490.65 & 0.31 & 3.35 & 0.2708 \\
  N & GRB121001A & 373.95 & 335.69 & 421.89 & 0.23 & 0.81 & 0.4449 \\
  N & GRB121011A & 4044.76 & 3792.69 & 4241.53 & 0.11 & 1.42 & 0.3371 \\
  N & GRB121011A & 4505.02 & 4241.53 & 4747.37 & 0.11 & 1.48 & 0.3131 \\
  N & GRB121024A & 205.13 & 182.87 & 247.53 & 0.32 & 0.31 & 0.9336 \\
  N & GRB121024A & 277.11 & 247.53 & 352.16 & 0.38 & 2.08 & 0.6950 \\
  N & GRB121027A & 6074.90 & 1150.72 & 35420.68 & 5.64 & 442.82 & 0.9853 \\
  N & GRB121027A & 247.49 & 221.07 & 521.32 & 1.21 & 3.40 & 0.8134 \\
  N & GRB121028A & 765.23 & 620.09 & 1257.88 & 0.83 & 9.71 & 0.8283 \\
  N & GRB121102A & 56.96 & 54.23 & 61.32 & 0.12 & 0.51 & 0.7088 \\
  N & GRB121108A & 139.66 & 104.34 & 503.56 & 2.86 & 113.53 & 1.0000 \\
  N & GRB121108A & 623.81 & 751.62 & 823.91 & 0.12 & -0.20 & 0.4341 \\
  N & GRB121117A & 83.17 & 76.32 & 104.76 & 0.34 & 0.40 & 0.5248 \\
  N & GRB121123A & 244.22 & 190.70 & 917.74 & 2.98 & 18.76 & 1.0000 \\
  N & GRB121125A & 91.89 & 88.06 & 122.54 & 0.38 & 1.60 & 0.9352 \\
  N & GRB121128A & 91.35 & 88.11 & 127.72 & 0.43 & 0.67 & 0.8100 \\
  N & GRB121209A & 80.29 & 80.29 & 133.72 & 0.67 & 8.88 & 0.9983 \\
  N & GRB121211A & 175.84 & 124.51 & 294.51 & 0.97 & 9.78 & 1.0000 \\
  N & GRB121211A & 96.63 & 85.88 & 105.46 & 0.20 & 1.25 & 0.5686 \\
  N & GRB121212A & 221.09 & 131.29 & 494.36 & 1.64 & 16.57 & 0.9999 \\
  N & GRB121212A & 577.29 & 494.36 & 939.07 & 0.77 & 17.14 & 0.8500 \\
  N & GRB121212A & 58.48 & 58.48 & 131.29 & 1.25 & 3.96 & 0.6032 \\
  N & GRB121217A & 736.16 & 222.19 & 1625.91 & 1.91 & 165.87 & 1.0000 \\
  N & GRB121226A & 197.74 & 140.95 & 275.39 & 0.68 & 1.30 & 0.8769 \\
  N & GRB121229A & 458.62 & 344.76 & 5173.39 & 10.53 & 35.27 & 1.0000 \\
  N & GRB121229A & 232.48 & 216.72 & 261.15 & 0.19 & 0.51 & 0.4989 \\

\enddata
\tablecomments{Flares are listed in chronological order by GRB date, then sorted by confidence.  *All times are relative to the time of the initial burst trigger.  $\Delta t/t$ is calculated as ($T_{stop} - T_{start})/T_{peak}$.  $T_{start}$ and $T_{stop}$ are lower and upper limits, respectively.  Flux Ratio is calculated as the flux at the flare peak time divided by the extrapolated flux of the underlying light curve at the same time, normalized using the flux of the underlying light curve, and is a lower limit of the actual peak flux ratio.  The confidence measure represents the fraction of times the flare was identified during the 10,000 Monte Carlo simulations.  The first column identifies whether the identified feature comes from an overlapping `fairing period'.}
\label{tab:Flaretable}
\end{deluxetable*}
\clearpage

\end{document}